\documentstyle[12pt]{article}
\def\abstract#1{\vskip 7mm 
	\begin{center}{\large Abstract}\par \bigskip
		\begin{minipage}[c]{12cm}
			\small #1
		\end{minipage}
	\end{center}
}
\def\title#1{\begin{center}{\Large\bf #1}\end{center}}
\def\author#1{\vskip 5mm \begin{center}{#1}\end{center}}
\def\address#1{\begin{center}{\it #1}\end{center}}

\newcommand{\bfr}{\begin{flushright}}
\newcommand{\efr}{\end{flushright}}

\begin{document}

\vspace*{-2cm}
\bfr{}\efr\vspace{-9mm}
\bfr{}\efr
\vspace{1cm}

\title{Equivalence of Weyl Vacuum and Normal Ordered Vacuum in the Moyal Quantization}
\author{Takao KOIKAWA\footnote{E-mail: koikawa@otsuma.ac.jp}}
\vspace{1cm}
\address{
  School of Social Information Studies,
         Otsuma Women's University\\
         Tama 206-0035,~Japan\\
}
\vspace{2.5cm}

We study the features of the vacuum of the harmonic oscillator in the Moyal
quantization. Two vacua are defined, one with the normal ordering and the other with the Weyl ordering. Their equivalence is shown by using a differential equation satisfied by the normal ordered vacuum.

\newpage
\setcounter{page}{2}
There has been a great interest in another way of the quantization, or the  Moyal quantization~\cite{Moy}, especially in its relation to the super string theory.
In the previous paper~\cite{Koi}, we studied the vacuum of the harmonic oscillator in the Moyal quantization. We can introduce a pair of new variables $a$ and $a^{\dagger}$ instead of the phase variables $q$ and $p$, which satisfy the Moyal bracket commutation relation isomoprhic to the operator commutation relation(CR hereafter) between the creation and the annihilation operators. The Hamiltonian is expressed in terms of $a$ and $a^\dagger$ with the star product between them. Once the vacuum is obtained, the $\star$-gen function belonging to the $\star$-gen value $(n+\frac{1}{2})\hbar$ can be obtained by successively multiplying $a^\dagger$ from the left and $a$ from the right n times to the vacuum~\cite{Curt,Fair}.
In the Moyal quantization, the lowest energy level solution, which can be expressed in terms of the Weyl products of $a$ and $a^\dagger$ is natural to be called the vacuum because it vanishes when it is multiplied by $a$ from the left or by $a^\dagger$ from the right. We call it the Weyl vacuum. There is another way of defining the vacuum. On the analogy to the operator formalism, we can introduce the normal ordering of the star products. Using this, we can also define the vacuum that we call the normal ordered vacuum.
The apparent difference between two vacua raises a question. Are they identical or not? In the previous paper, we expanded two vacua in terms of $1/\hbar$ and compared them. When we reorder the Weyl ordered vacuum into the normal ordered one, there appear the infinities at lower order of $1/\hbar$. As long as we use the expansion in terms of $1/\hbar$ for the comparison, what we get is at most the result order by order.  Then, it is difficult to conclude anything definitive.

Here in this paper, we study the same subject from a different point of view. Instead of rewriting the vacuum function in terms of the star products of $a$ and $a^\dagger$ as in the previous paper, we try to rewrite the normal ordered vacuum into a function explicitly and then compare two vacua. In order to obtain the functional form of the normal ordered vacuum, we derive a differential equation satisfied by a class of functions including the normal ordered vacuum, and then solve it. The solution includes the integral constants together with a parameter $t$ which is introduced to obtain the differential equation. These integral constants are determined by the boundary condition at $t=0$. The value of the parameter is set so that the function coincides with the normal ordered vacuum. We thus determine the functional form of the normal ordered vacuum. 

First we give the definitions and notation of the Moyal star product. The star product of two functions $f=f(q,p)$ and $g=g(q,p)$ is defined by
\begin{equation}
f\star g = \exp \Bigg[ \hbar
\Bigg(\frac{\partial~}{\partial q}\frac{\partial~}{\partial\tilde p}-\frac{
\partial~}{\partial p}\frac{\partial~}{\partial\tilde q}\Bigg)
\Bigg] f({\bf x}) g({\bf \tilde x}) \vert_{{\bf x} = {\bf\tilde x}},
\label{eq:star1}
\end{equation}
where ${{\bf x}=(q,p)}$ and ${{\bf \tilde x}=(\tilde q,\tilde p)}$  and they are set equal after the derivatives are taken.

In the analysis of the harmonic oscillator, it is more convenient to use the complex variable $z$ and its complex conjugate variable $\bar z$ instead of the phase variables $q$ and $p$. We define $z$ as $z=q+ip$.
We list the formulas for these variables:
\begin{eqnarray}
z \star f(z,\bar z)&=&\Big(z+\hbar \frac{\partial}{\partial \bar z}\Big) f(z,\bar z),
\label{eq:zfrml1}\\
\bar z \star f(z,\bar z)&=&\Big(\bar z-\hbar \frac{\partial}{\partial  z}\Big)f(z,\bar z),
\label{eq:zfrml2}\\
f(z,\bar z) \star z&=&\Big(z-\hbar \frac{\partial}{\partial \bar z}\Big)f(z,\bar z),\label{eq:zfrml3}\\
f(z,\bar z) \star {\bar z}&=&\Big({\bar z}+\hbar \frac{\partial}{\partial  z}\Big)f(z,\bar z).
\label{eq:zfrml4}
\end{eqnarray}
We denote the star product CR of $f(z,\bar z)$ and $g(z,\bar z)$
as
\begin{equation}
[f(z,\bar z),g(z,\bar z)]= f(z,\bar z)\star g(z,\bar z)-g(z,\bar z)\star f(z,\bar z).
\end{equation}

Using the above formulas, we obtain the star product CR of $z$ and $\bar z$:
\begin{equation}
[z,\bar z]=2\hbar.
\label{eq:zcr}
\end{equation}
By introducing the new variables $a$ and $a^{\dagger}$ by $a=z/{\sqrt 2}$ and $a^\dagger=\bar z/{\sqrt 2}$,
the star product CR reads
\begin{equation}
[a,a^\dagger]=\hbar,
\label{eq:crofa}
\end{equation}
which shows that the star product CR of $a$ and $a^\dagger$ is isomorphic to  the operator CR of the creation and annihilation operators. 

The Hamiltonian that we consider is that of the harmonic oscillator,
\begin{equation}
H=\frac{p^2}{2m}+k\frac{q^2}{2},
\end{equation}
where we assume $m=k=1$. This can be rewritten using the variables $z$ and $\bar z$ or $a$ and $a^\dagger$ as
\begin{equation}
H=\frac{1}{2}{\bar z}z=\frac{1}{2}({\bar z}\star z+\hbar)=a^\dagger \star a+\frac{\hbar}{2}.
\label{eq:Ham}
\end{equation}

We obtain the $\star$-gen function $f_n$ belonging to the eigenvalue $E_n$ by solving the $\star$-gen value equation~\cite{Fair}
\begin{equation}
H(z,\bar z) \star f_n(z,\bar z)=E_nf_n(z,\bar z).
\label{eq:stareq}
\end{equation}
The $\star$-gen function $f_n$ belonging to the energy level $(n+\frac{1}{2})\hbar$(n=0,1,2$\cdots$)   is expressed by the Laguerre polynomial $L_n(x)$ as
\begin{equation}
f_n(H)=e^{-\frac{2H}{\hbar}}L_n(2H),(n=0,1,2,\cdots),
\end{equation}
where $H=(q^2+p^2)/2$.
The 0-th $\star$-gen function $f_0$ belonging to the lowest level is
\begin{equation}
f_0={e}^{-\frac{2H}{\hbar}}={e}^{-\frac{\bar z z}{\hbar}},
\label{eq:wvac}
\end{equation} 
which can be regarded as the vacuum. Other states are constructed by operating with $\bar z$ from the left and $z$ from the right repeatedly. 

When (\ref{eq:wvac}) is expanded, each term $(\bar z z)^n$ is expressed by using the Weyl ordering of the star products. The Weyl ordering was first introduced as a map from classical functions to operator products in the quantization. Here we discuss the Weyl ordering of the star products within the classical framework. We denote the sum of the possible permutations of the star products of $n$ variables $\bar z$ and $n$ variables $z$ by Perm($\bar z^n$, $z^n$). Dividing this by the number of terms, we define the Weyl ordering of the $n$ variables $\bar z$ and $n$ variables $z$ by 
\begin{equation}
(\bar z^n z^n)_W=\frac{(n!)^2}{(2n)!} {\rm{Perm}}(\bar z^n,z^n).
\end{equation}
One of the simplest examples is in the case $n=1$:
\begin{equation}
(\bar z z)_W=\frac{1}{2}(\bar z \star z+z \star \bar z).
\end{equation}
In general, we can express the power of $\bar z z$ in terms of the  Weyl ordered star products as 
\begin{equation}
(\bar z z)^n=(\bar z^n z^n)_W.
\end{equation}
We are thus able to express the vacuum in terms of the Weyl ordered star products. We call this the Weyl vacuum.

We can also consider another type of vacuum by introducing the normal ordering of the star products. This is given by
\begin{equation} 
:e^{-\frac{1}{2\hbar}\bar z \star z}:=:e^{-\frac{a^{\dagger}\star a}{\hbar}}:\sim:e^{-\frac{H}{\hbar}}:, 
\end{equation} 
where the double dots denote the normal ordering of the star products which is defined, in analogy to the operator case, by putting all $a^\dagger$ to the left of all $a$. We call this the normal ordered vacuum. Explicitly they are written as
\begin{equation}
:e^{-\frac{a^\dagger \star a}{\hbar}}:
=1+(-\frac{1}{\hbar})a^\dagger \star a+\frac{1}{2!}(-\frac{1}{\hbar})^2 
a^\dagger \star a^\dagger \star a \star a+\cdots.
\label{eq:nvac}
\end{equation}

We now have the Weyl vacuum and the normal ordered vacuum. These two vacua do not look alike, which motivates the present and the previous study.
The imaginary part of Eq.(\ref{eq:stareq}) requires that the $\star$-gen function should be a function of the variable $H$. The function $\phi(H)$ representing the vacuum in the Moyal quantization is characterized by the condition
\begin{equation} 
a\star \phi(H)=\phi(H)\star {a^\dagger}=0.
\label{eq:defvac}
\end{equation}
When the vacuum is obtained, the functions belonging to the higher levels are obtained algebraically by making use of the star product CR (\ref{eq:crofa}) and the expression of the Hamiltonian in terms of $a$ and $a^\dagger$ in (\ref{eq:Ham}). It is easy to show that both vacua satisfy this vacuum condition. 
 
The Weyl vacuum has an explicit function form, while the normal ordered vacuum has an operator-like expression. Then, there can be two standpoints in the comparison of two vacua. One is to rewrite the function into an operator-like form by introducing the Weyl ordering, which we did in the previous paper~\cite{Koi}. The other is to rewrite the operator-like expression into a function, which we discuss here.
Introducing a parameter $t$, we define $f=f(t,H)$ by 
\begin{equation}
f(t,H)=:{\rm e}^{t{\bar z}\star z}:,
\end{equation}
which satisfies the condition
\begin{equation}
f(0,H)=1.
\label{eq:cond}
\end{equation}
We also note that, when the parameter $t$ is set to be $-(1/2\hbar)$, it becomes identical to the normal ordered vacuum:
\begin{equation}
f(-\frac{1}{2\hbar},H)=:e^{-\frac{a^\dagger \star a}{\hbar}}:.
\label{eq:cnsstnt}
\end{equation}
Differentiating $f(t,H)$ with respect to the variable $t$, we obtain
\begin{equation}
\frac{\partial f(t,H)}{\partial t}={\bar z} \star f(t,H) \star z.
\end{equation}
Using the formulas (\ref{eq:zfrml2}) and (\ref{eq:zfrml3}), we rewrite the
 equation as
\begin{equation}
\frac{\partial f}{\partial t}=
2\Big\{ Hf+\Big( \frac{\hbar}{2}\Big)^2(f'+Hf'')-\Big( \frac{\hbar}{2}\Big)(f+2Hf')\Big\},
\label{eq:difeq}
\end{equation}
where $f'=\frac{\partial f}{\partial H}$ and  $f''=\frac{\partial^2 f}{\partial H^2}$.

The solution is given by
\begin{equation}
f(t,H)=\frac{C_2}{C_1-2t}{\rm e}^{\Big( \frac{2}{\hbar}\Big)\Big(1+\Big( \frac{2}{\hbar}\Big)\frac{1}{C_1-2t}\Big)H},
\label{eq:ff}
\end{equation}
where $C_1$ and $C_2$ are the integral constants. 
These constants are determined by using the condition (\ref{eq:cond}) as
\begin{eqnarray}
1+\Big( \frac{2}{\hbar}\Big)\frac{1}{C_1}&=&0,\\
\frac{C_2}{C_1}&=&1.
\end{eqnarray}
We thus obtain
\begin{equation}
C_1=C_2=-\frac{2}{\hbar}.
\end{equation}
Substituting these results into (\ref{eq:ff}), we obtain $f$ as
\begin{equation}
f(t,H)=\frac{1}{1+\hbar t}{\rm e}^{\frac{2t}{1+\hbar t}H}.
\end{equation}
Setting $t=-(1/2\hbar)$, which reduces $f$ to the normal ordering vacuum as we mentioned in (\ref{eq:cnsstnt}), we obtain
\begin{equation}
f(-\frac{1}{2\hbar},H)=2{\rm e}^{-\frac{2H}{\hbar}},
\label{eq:sol}
\end{equation}
which is identical to the Weyl vacuum.

In order to show the uniqueness of the solution, we note that the equation (\ref{eq:difeq}) is linear in $f(t,H)$, and so any linear combination of the solutions would be the solution to the equation. Suppose that there is a solution $g(t,H)$ which is assumed to be different from the solution (\ref{eq:sol}). We further require that it should satisfy the vacuum conditions (\ref{eq:defvac}).
By using the formulas (\ref{eq:zfrml1}) and (\ref{eq:zfrml4}), they are rewritten as
\begin{equation}
zg(t,H)+\hbar \frac{\partial g(t,H)}{\partial {\bar z}}={\bar z}g(t,H)-\hbar \frac{\partial g(t,H)}{\partial z}=0.
\end{equation}
These lead to a differential equation for $g(t,H)$:
\begin{equation}
\frac{\hbar}{2}\frac{\partial g(t,H)}{\partial H}+g(t,H)=0,
\end{equation}
which is solved to yield
\begin{equation}
g(t,H)\sim {\rm e}^{-H/{2\hbar}}.
\end{equation}
The $H$ dependence of this solution is the same as the solution (\ref{eq:sol}). Therefore
this contradicts the assumption that $g(t,H)$ is different from $f(t,H)$.
We thus conclude that the solution (\ref{eq:sol}) is a unique vacuum solution to Eq.(\ref{eq:difeq}), and so the normal ordered vacuum is equivalent to the Weyl ordered vacuum up to an overall constant factor. 

When we estimated the Weyl ordered vacuum by expanding it in terms of $1/\hbar$ and reordering it into the normal ordered form, there appear infinities at lower order of $1/\hbar$~\cite{Koi}. By taking the present result, or the equivalence of two vacua, into account, the appearance of infinities suggests that they are factored out to be an overall constant when all of them are summed up, which elucidates what the renormalization is like in the Moyal quantization.  

\vspace{2cm}

The author would like to acknowledge the Doyou-kai members for useful discussion.

\end{document}